\documentclass[fleqn,usenatbib]{mnras}

\usepackage[T1]{fontenc}
\usepackage{ae,aecompl}

\DeclareRobustCommand{\VAN}[3]{#2}
\let\VANthebibliography\thebibliography
\def\thebibliography{\DeclareRobustCommand{\VAN}[3]{##3}\VANthebibliography}

\usepackage{graphicx}	
\usepackage{amsmath}	
\usepackage{amssymb}	
\usepackage{xcolor}
\usepackage{lastpage}

\newcommand{\msun}{\mathrm{M}_\odot}

\newcommand{\cmspg}{\,\mathrm{cm}^2 \mathrm{g}^{-1}}

\newcommand{\kb}{k_\mathrm{B}}
\newcommand{\dg}{^\circ}
\newcommand{\ye}{Y_\mathrm{e}}

\makeatletter
\renewcommand\d[1]{\mspace{6mu}\mathrm{d}#1\@ifnextchar\d{\mspace{-3mu}}{}}
\makeatother

\usepackage{newtxtext,newtxmath}

\title[Dynamical r-process heating in disc winds]{The impact of r-process heating on the dynamics of neutron star merger accretion disc winds and their electromagnetic radiation}

\author[H. Klion et al.]{
Hannah Klion$^{1,2}$\thanks{E-mail: hklion@berkeley.edu},
Alexander Tchekhovskoy$^{3}$,
Daniel Kasen$^{1,2,4}$,
Adithan Kathirgamaraju$^{1}$,
\newauthor Eliot Quataert$^{5}$, and
Rodrigo Fern\'{a}ndez$^{6}$
\\
$^{1}$Astronomy Department and Theoretical Astrophysics Center, University of California, Berkeley, Berkeley, CA 94720, USA\\
$^{2}$Physics Department, University of California, Berkeley, Berkeley, CA 94720, USA\\
$^{3}$Center for Interdisciplinary Exploration \& Research in Astrophysics (CIERA), Physics \& Astronomy, Northwestern University, Evanston, IL 60208, USA\\
$^{4}$Nuclear Science Division, Lawrence Berkeley National Laboratory, Berkeley, CA 94720, USA\\
$^{5}$Department of Astrophysical Sciences, Princeton University, Princeton, NJ 08544, USA\\
$^{6}$Department of Physics, University of Alberta, Edmonton, AB T6G 2E1, Canada
}

\date{Accepted XXX. Received YYY; in original form ZZZ}

\pubyear{2021}
\begin{document}
\label{firstpage}
\pagerange{\pageref{firstpage}--\pageref{lastpage}}
\maketitle

\begin{abstract}
Neutron star merger accretion discs can launch neutron-rich winds of $>10^{-2} \msun$. This ejecta is a prime site for r-process nucleosynthesis, which will produce a range of radioactive heavy nuclei. The decay of these nuclei releases enough energy to accelerate portions of the wind by $\sim 0.1c$. 
Here, we investigate the effect of r-process heating on the dynamical evolution of disc winds. We extract the wind from a 3D general relativistic magnetohydrodynamic simulation of a disc from a post-merger system. This is used to create inner boundary conditions for 2D hydrodynamic simulations that continue the original 3D simulation. We perform two such simulations: one that includes the r-process heating, and another one that does not.
We follow the hydrodynamic simulations until the winds reach homology (60 seconds). Using time-dependent multi-frequency multi-dimensional Monte Carlo radiation transport simulations, we then calculate the kilonova light curves from the winds with and without dynamical r-process heating. We find that the r-process heating can substantially alter the velocity distribution of the wind, shifting the mass-weighted median velocity from $0.06c$ to $0.12c$. The inclusion of the dynamical r-process heating makes the light curve brighter and bluer at $\sim 1\,\mathrm{d}$ post-merger. However, the high-velocity tail of the ejecta distribution and the early light curves are largely unaffected.
\end{abstract}




\section{Introduction}
GW170817 was the first gravitational wave detection of a binary neutron star (NS) merger \citep{abbott:17_nsgw}. It was followed by a short gamma ray burst (sGRB), confirming the connection between compact object mergers and sGRBs \citep{abbott:17_grgw, goldstein:17, savchenko:17}. An ultraviolet, optical, and infrared counterpart, AT 2017gfo (also SSS17a), was detected hours later, consistent with a $v\sim 0.1c$ outflow of $>0.01\msun$ of radioactive material with an average opacity of $1-3\cmspg$ \citep{abbott:17_multimessenger, arcavi:17,kasliwal:17, drout:17, villar:17, coulter:17, cowperthwaite:17, chornock:17, soares-santos:17}. This was consistent with predictions of a ``kilonova'', a radioactively-powered, red, and rapidly evolving counterpart of a NS merger \citep{metzger:10, metzger:12, barnes:13, tanaka:13}.

Outflows from NS mergers are expected via multiple channels.
As an NS binary merges, mass ejection occurs on dynamical $(\sim \mathrm{ms})$ time-scales due to hydrodynamical and tidal forces. Numerical simulations predict that dynamical ejecta consist of $10^{-4} - 10^{-2} \msun$ of material with escape velocities of $\sim 0.1-0.3c$ \citep{bauswein:13, hotokezaka:13_massej, lehner:16,  bovard:17, dietrich:17, radice:18_dynej_eos}. Some of the disrupted material is still gravitationally bound and can form an accretion disc of up to $0.3 \msun$ that evolves on longer timescales ($100\,\mathrm{ms}-10\,\mathrm{s}$) \citep{oechslin:07, hotokezaka:13_mnsgw}. Initially, neutrino cooling dominates because the disc is hot and dense \citep{popham:99, narayan:01}. As the disc cools and spreads, neutrino cooling becomes inefficient, and the disc becomes fully advective. Weak interactions freeze out, which can lead to a strong neutron-rich wind \citep{metzger:09}. 

Regardless of how it is launched, the ejected material undergoes rapid decompression from nuclear densities. Once the material cools below $\sim 10\,\mathrm{GK}$ and leaves nuclear statistical equilibrium (NSE), neutrons capture onto light seed nuclei faster than the nuclei can undergo beta decays. This rapid- (r-)process nucleosynthesis produces a range of radioactive neutron-rich nuclei that beta decay to stability over weeks \citep{lattimer:77, eichler:89, metzger:10}. The bulk of the heat released from r-process decays is deposited in the first few seconds after the material leaves NSE. It then falls off approximately as a power law, powering the kilonova \citep{li:98, metzger:10}. 

The distribution of nuclei produced by the r-process primarily depends on the electron fraction $\ye$ of the nuclear material. When $\ye \lesssim 0.25$, the r-process produces material enriched with Lanthanides and Actinides, which have uniquely high opacities. Their partially-filled $f$ orbitals produce a high density of moderately strong atomic lines, which lead to a high quasi-continuum opacity ($\kappa \sim 10 \cmspg$), especially in the blue and ultraviolet \citep{kasen:13, fontes:15, tanaka:20}. Low $\ye$ ejecta would therefore produce a red, dim, slowly-evolving transient \citep{barnes:13, wollaeger:18}. By contrast, higher-$\ye$ material gives rise to lighter r-process elements, which primarily occupy the second row of the $d$-block. These elements have lower average opacities $\kappa \sim 0.5-1\cmspg$ that are again highly wavelength-dependent.
The light curve of AT 2017gfo can be modeled by (at least) two distinct components with opacities that roughly correspond to light and heavy r-process products \citep{cowperthwaite:17, kasen:17, perego:17, villar:17}.

Many past studies of post-merger accretion discs have relied on axisymmetric hydrodynamic simulations, where an imposed shear viscosity transports angular momentum. \citep[e.g.][]{fernandez:13, just:15, fujibayashi:18}.
\citet{siegel:17, siegel:18} presented the first 3D general relativistic magnetohydrodynamic (GRMHD) simulations of the post-merger accretion disc evolution. They track the system out to 400 ms, by which point about half of the wind has been launched. \citet[hereafter F19]{fernandez:19} followed the evolution of a post-merger disc system to 9 seconds, allowing all of the disc material to be accreted or ejected.
\citet{miller:19_diskblue} evolve the wind to $130$ ms with concurrent GR Monte Carlo neutrino transport, and with a full nuclear reaction network. They find that the resulting outflow primarily consists of light r-process elements, consistent with a blue disc wind. The wind is also sensitive to the initial magnetic field configuration within the disc; toroidal or weaker fields lead to less massive and slower outflows \citep{christie:19}.

The very long-term evolution of disc winds remains uncertain.
Most disc wind simulations are only evolved to $\lesssim 10\,\mathrm{s}$, while the wind needs to expand for several times that in order to reach homology, which is required to generate light curves and spectra with most existing photon radiative transfer codes. Longer-term evolution of hydrodynamic winds has been studied by extracting wind properties at a large radius, and using that to set an inner boundary condition on a larger grid \citep[e.g.][]{kasen:15, kawaguchi:21}. This technique has not yet been applied to a GRMHD simulation of a disc wind. 

R-process heating provides around $1$--$3\,\mathrm{MeV}$ per nucleon, most of which is deposited within 1 s. If this were completely converted to kinetic energy, it would boost a particle at rest to a velocity of $0.1-0.2 c$. Prior work has found that r-process heating can have a particularly strong effect on the tidal component of the dynamical ejecta, causing it to inflate and smooth out inhomogeneities \citep{rosswog:14, grossman:14, fernandez:15_dyndisc, darbha:21}.

In this paper, we are interested in understanding how r-process heating affects the structure of a 3D MHD-driven disc wind and, consequently, the resulting kilonova light curves. We evolve a kilonova disc wind to homology both with and without the r-process heating. We construct an inner boundary condition for 2D GR hydrodynamics (GRHD) simulations based on the wind formed in the 3D GRMHD simulation of F19, approximately accounting for r-process heating that occurs before injection. 
We present our formalism for doing so in Section \ref{harmsedona:sec:method_harm_ic}, and discuss the details of our 2D GRHD simulations in Section \ref{harmsedona:sec:method_harm}. After $60\, \mathrm{s}$ of evolution, we pass the resulting density and temperature structures to \textsc{Sedona}, a multi-dimensional, multi-frequency radiation transport code (Sections \ref{harmsedona:sec:method_sedona_ic} and \ref{harmsedona:sec:method_sedona}). In Section \ref{harmsedona:sec:result_harm}, we compare the results of our GRHD simulations with and without r-process heating to assess the effect of r-process heating on the dynamical evolution of kilonova disc winds. Our predictions for disc wind light curves, with and without dynamic r-process heating, are found in Section \ref{harmsedona:sec:result_sedona}. We conclude and discuss future research directions in Section \ref{harmsedona:sec:discussion}. 

\section{Methods}
\label{harmsedona:sec:methods}

Our full calculation consists of the following steps: 
\begin{enumerate}
    \item Simulate a post-merger black hole accretion disc in 3D GRMHD to $10\,\mathrm{s}$. This model is the one presented in F19, and does not include r-process heating.
    \item Extract wind properties from F19 at a radius of $r_b = 2\times 10^4\,\mathrm{km}$, and use them to set an inner boundary condition for a 2D GRHD simulation. There are two versions of these conditions: one that is simply axisymmetrized, and another that is axisymmetrized and then modified to approximately account for r-process heating that occurred within the domain of the 3D simulation, prior to the gas reaching $r_b$.
    \item Evolve two 2D GRHD simulations, each with one of the sets of boundary conditions described above. The simulation `with r-process' uses the r-process-adjusted inner boundary conditions and self-consistently includes an r-process heating source term during evolution. Meanwhile, the simulation `without r-process' does not include an r-process heating term, and uses the axisymmetrized boundary conditions (not modified to include r-process heating). We evolve both of these models until $60\,\mathrm{s}$, at which point they are mostly in free expansion.
    \item Calculate the optical emission from both of the 2D GRHD simulations using \textsc{Sedona}, a Monte Carlo radiation transport code. In this phase of the calculation, both models include r-process heating, but the evolution is assumed to be homologous. The velocity structure is fixed, so the heating only affects the light curve. This allows us to isolate the effects of r-process heating on the dynamics of the ejecta.
\end{enumerate}

\subsection{2D GR hydrodynamic initial conditions}
\label{harmsedona:sec:method_harm_ic}

We continue the 3D GRMHD simulation of F19 in 2D GRHD. Unless otherwise stated, we take $G=c=1$. F19 initialize a torus of mass $0.033\msun$ \footnote{The simulations of F19 were performed before the announcement of GW170817/AT 2017gfo. Subsequent modeling has inferred that the initial torus mass was larger by a factor of $\sim 3$ \citep{shibata:17}.} with a strong poloidal magnetic field around a black hole of mass $3\msun$ and  spin parameter $a=0.8$.
 We extract the time-series of the primitive variables at a radius of $r_b \equiv 2 \times 10^4\,\mathrm{km} = 4452 \,r_g$ in the 3D simulations: rest mass density $\rho$, energy density $\varepsilon$, four velocity in Kerr-Schild coordinates $\{u^t, u^r, u^\theta, u^\phi\}$ and composition (electron fraction $\ye$ and abundances of protons, neutrons, and alpha particles). This radius is small enough that the entire wind crosses through the surface before the end of the 3D simulation, but large enough that no bound material falls back through.
 
 We make the simplifying assumption that the magnetic fields are zero. While they are critical for the jet launch and wind ejection, both of these occur interior to the inner boundary of our simulation. We are primarily interested in the baryon-rich wind, and do not expect the exclusion of magnetic fields to significantly affect our results.

We axisymmetrize the density by averaging in $\phi$:
\begin{equation}
    \langle \rho\rangle = \frac{\int \rho \sqrt{-g} \d \phi}{\int \sqrt{-g} \d \phi},
\end{equation}
and calculating mass-weighted averages for the other variables, $X$:
\begin{equation}
    \langle X \rangle = \frac{\int X \rho \sqrt{-g} \d \phi}{\int \rho \sqrt{-g} \d \phi}.
\end{equation}
The determinant of the metric is $g$. These axisymmetrized values are then used as a time-dependent inner boundary condition in 2D GRHD simulations. We perform two simulations: one with and one without r-process heating. When we do not include r-process heating, we directly use the axisymmetrized primitives at $r_b$ as the boundary condition. 

\begin{figure*}
  \includegraphics{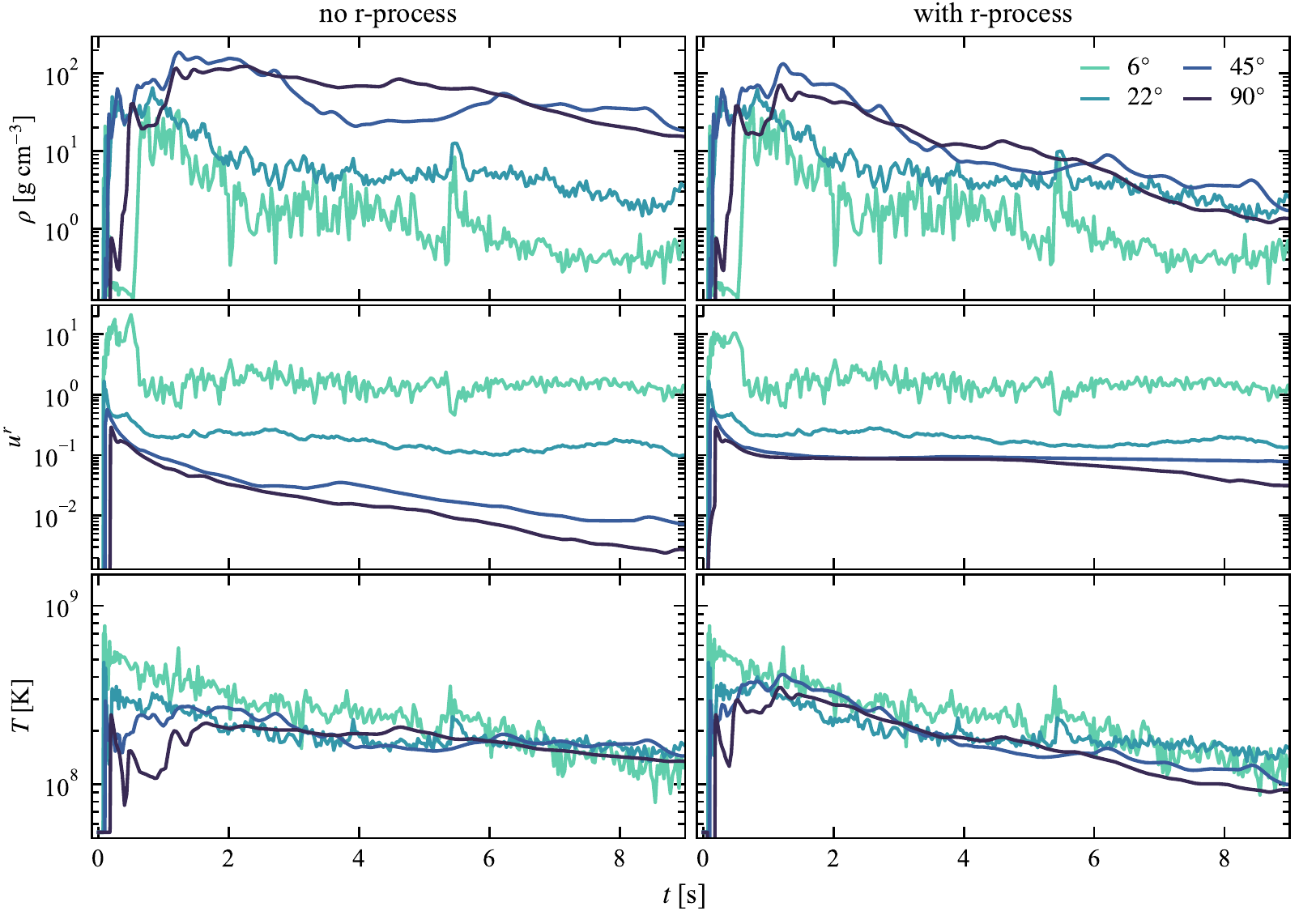}
  \caption{Inner boundary condition at $r_b = 2\times 10^4\,\mathrm{km}$ in the 2D GRHD simulations of long-term disc wind evolution (see Section \ref{harmsedona:sec:method_harm_ic}), derived from the 3D GRMHD simulation of F19. The left (right) column shows the conditions used for the simulation without (with)  r-process heating. The conditions on the right include approximate effects of r-process heating that occurs before injection. We show rest mass density (top panel), radial four-velocity (middle), and radiation temperature (bottom) as a function of time for a representative selection of angles. Temperature is calculated from internal energy density $\varepsilon$ assuming radiation pressure dominates, $T = (\varepsilon/a)^{1/4}$, where $a$ is the radiation constant. 
  The conditions are largely symmetric across the equator, so we only show angles in the Southern hemisphere; angles are measured from the South pole. The effects of r-process heating are more pronounced toward the higher-density, lower-velocity equatorial region.}
    \label{harmsedona:fig:inflow_lines}
\end{figure*}

\begin{figure}
    \centering
    \includegraphics{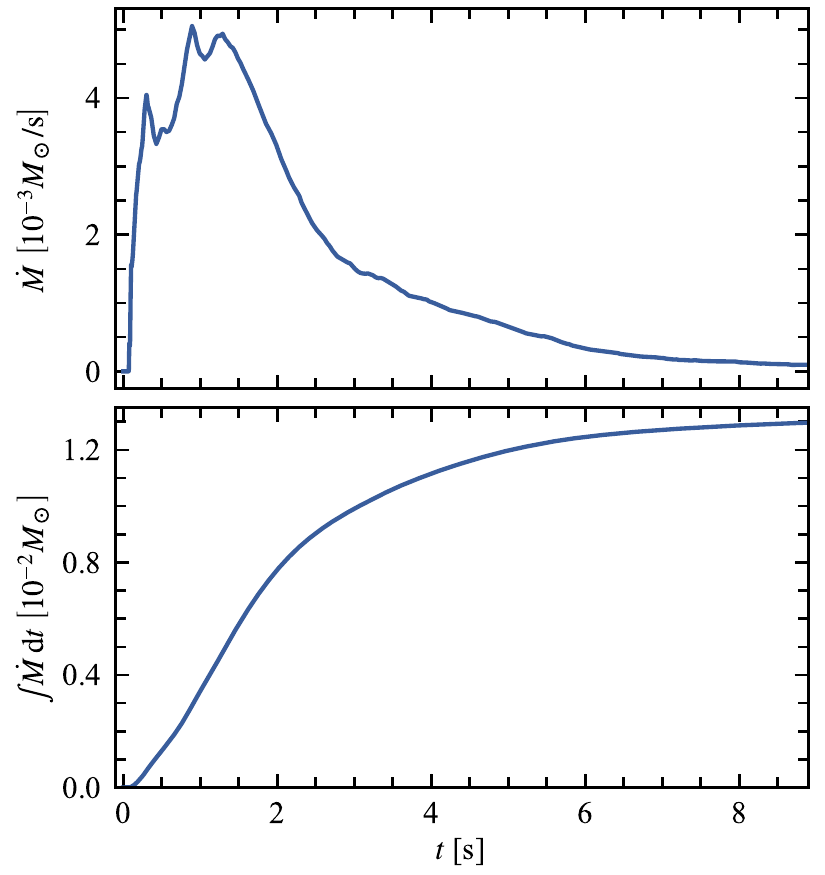}
    \caption{Mass injection rate $\dot M$ (top panel) and total injected mass (bottom panel) in our GRHD simulations as a function of time. The majority of the mass injection occurs within the first few seconds of the simulation, with a median injection time of 1.6 s. By construction, the mass injection rate is the same in the HARM runs with and without r-process heating.}
    \label{harmsedona:fig:mdot_in}
\end{figure}

The majority of the thermal energy input by the r-process will be converted to kinetic energy via adiabatic expansion of the hot gas. The decay of r-process elements deposits $\sim 4 \times 10^{18}\,\mathrm{erg\,g}^{-1}$ on a timescale of $\sim 1 \,\mathrm{s}$, which corresponds to a velocity increase of $\sim 0.1c$ for a particle at rest. Of the mass crossing $r_b$ in F19, $8.7 \times 10^{-3} \msun$ (68 per cent) has a velocity $<0.1c$, so we expect acceleration from this burst of energy to be significant. The effects of $\ye$ on heating rates are generally modest \citep{lippuner:15}, so we assume a spatially uniform, time-dependent heating rate. We adopt the $\ye=0.1$ r-process heating rate from \citet{metzger:10}. We approximate the time-dependent heating rate per mass $\eta(t)$ as a broken power law:
\begin{equation}
\label{harmsedona:eq:erad}
    \eta(t) = \begin{cases} 
    6.7\times 10^{17}\,\mathrm{erg/g/s} & t_1 < 0.01,\\[6pt]
    1.1\times10^{19} t_1^{0.6}\,\mathrm{erg/g/s} & 0.01 \leq t_1 < 0.5,\\[6pt]
    7.0\times 10^{17} t_1^{-3.33}\,\mathrm{erg/g/s} & 0.5 \leq t_1 < 4,\\[6pt]
    4.2 \times 10^{16} t_1^{-1.3}\,\mathrm{erg/g/s} & t_1>4.
    \end{cases}
\end{equation}
where $t_1 = t/1\,\mathrm{s}$. This rate assumes that all of the nuclear energy emitted by radioactive decays is converted to thermal energy. In the first several seconds, neutrinos are the only decay product that can easily escape, while the remainder are thermalized. Neutrinos likely carry away $\sim 20-30$ per cent of the total energy \citep{metzger:10,barnes:16}. This adjustment would be less than the uncertainty in the r-process heating rate ($\sim$ factor of a few). We therefore assume complete thermalization during the hydrodynamic phase of the calculation.
We take $t=0$ for the r-process to be at the start of the simulation of F19. Physically, the r-process begins once the material leaves NSE, which will occur when the temperature drops below $\sim 10\,\mathrm{GK}$. These temperatures are only achieved within a radius of $50 r_g = 225\,\mathrm{km} \ll r_b$, and at early times $t\lesssim 300 \,\mathrm{ms}$. By contrast, the median $r_b$-crossing time is $1.7\,\mathrm{s}$. By the time the wind enters our domain, it has been out of NSE for some time, and it is a reasonable approximation that it left NSE at $t = 0$.

Before we axisymmetrize, we adjust the velocities and internal energy to include the effects of r-process heating. To calculate the modified boundary condition at time $t_b$, we consider a fluid parcel of mass $m$ that is crossing $r_b$ at time $t_b$. We assume that it is expanding homologously and has been doing so since the start of the 3D simulation. This is an approximation since the parcel will accelerate, so its evolution is not actually homologous. Also, it would have been ejected from the disc at some time after the start of the 3D calculation. None the less, this should give a reasonable estimate of the distribution of r-process energy into thermal and kinetic. Improving on this approximation would require the fluid trajectories, which we do not have. The parcel has volume $V(t)$ at time $t$. Unless otherwise indicated, we work in the fluid frame. The gas has been heated radioactively since $t_0=0\,\mathrm{s}$ by $\eta(t)$. Its thermal energy due to radioactive heating $e_\mathrm{rad}$ will evolve as
\begin{equation}
    \label{harmsedona:eq:rad_therm_diff}
    \frac{\d e_\mathrm{rad}}{\d t} = \frac{-e_\mathrm{rad}(t)}{t} + \eta(t) m.
\end{equation}
The first term on the right-hand side accounts for thermal losses due to adiabatic expansion. Solving the above differential equation and dividing by the volume $V(t)$ gives the internal energy density due to radioactive decays $\varepsilon_\mathrm{rad}$,
\begin{equation}
    \label{harmsedona:eq:rad_therm_int}
    \varepsilon_\mathrm{rad}(t) \equiv \frac{e_\mathrm{rad}(t)} {V(t)}= \frac{\rho(t)}{t} \int_{t_0}^{t} \eta(t') t' \d t'.
\end{equation}
We add $\varepsilon_\mathrm{rad}(t_b)$ to the internal energy density at $r_b$.
We assume that the remainder of the total r-process energy is converted to kinetic energy, giving the parcel a boost $\Delta e_\mathrm{kin}$:
\begin{equation}
    \label{harmsedona:eq:energy_tot}
    \Delta e_\mathrm{kin}(t_b) =  e_\mathrm{rad,tot}(t_b) - e_\mathrm{rad}(t_b),
\end{equation}
where
\begin{equation}
e_\mathrm{rad,tot}(t) = m \int_{t_0}^t \eta(t') \d t'
\end{equation}
is the total energy deposited by r-process heating up to time $t$.
The kinetic energy boost $\Delta e_\mathrm{kin} \geq 0$ because the thermal energy attributed to radioactive decays can never exceed the total heating by radioactivity. The conversion of non-radioactive thermal energy to kinetic energy is already handled in the underlying 3D simulation.

The energies and masses in this calculation are in the fluid frame. Ideally, we would use the fluid proper time to calculate the heating rate. Calculating the proper time for each parcel of fluid is infeasible since we do not have information about the trajectory of individual fluid elements. For the sake of consistency, we will use the observer time throughout this paper when evaluating $\eta(t)$. We do not expect this to have a substantial effect on our results. The vast majority of the mass has $v<0.2c$, so proper and observer time will generally be equal. Additionally, the bulk of the heating occurs in the first few seconds, so the total energy deposited over the  $\sim 10^2\,\mathrm{s}$ of the hydrodynamic simulation will be the same to within a few per cent. Effects on the dynamical evolution of the wind should therefore be negligible.

The combined rest mass and kinetic energy of the parcel is $\gamma m$, where $\gamma$ is the Lorentz factor. If we add $\Delta e_\mathrm{kin}$ in kinetic energy, the new energy of the parcel becomes
\begin{equation}
    \gamma_\mathrm{new} m = \gamma_\mathrm{old} m + \Delta e_\mathrm{kin}(t_b),
\end{equation}
and its new Lorentz factor is
\begin{equation}
\label{harmsedona:eq:quadenergy}
    \gamma_\mathrm{new} = \gamma_\mathrm{old} + \Delta e_\mathrm{kin}(t_b)/m.
\end{equation}

The magnitude of the new 3-velocity is $v_\mathrm{new}$, which can be calculated directly from $\gamma_\mathrm{new}$. We scale the components of the 3-velocity by $v_\mathrm{new} / v_\mathrm{old}$. While we set our boundary conditions in terms of densities, e.g.\ $\rho$ and $\varepsilon$, the physically significant quantity is the flux of mass and energy onto the grid. Accordingly, we scale the boundary $\rho$ and $\varepsilon$ by the ratio of the radial 4-velocities: $u^r_{\mathrm{old}} / u^r_{\mathrm{new}}$. This preserves the homologous estimate of the kinetic-thermal energy distribution.

Fig.~\ref{harmsedona:fig:inflow_lines} shows the time-dependent boundary condition used in our axisymmetric GRHD simulations for four representative angles. We show rest mass density, radial four-velocity ($u^r$) and temperature. We calculate the temperature from the internal energy density $\varepsilon$ assuming that the gas is radiation-dominated, $T = (\varepsilon/a)^{1/4}$, where $a$ is the radiation constant. Due to the jet, the polar regions have high velocity, particularly in the first $\sim 0.5\,\mathrm{s}$, where $u^r>1$. The high velocity in the polar regions is largely unaffected by the addition of r-process kinetic energy. However, the slower material in the bulk of the wind is accelerated to a velocity of $\sim 0.1c$. We hold the mass flux constant, so the density of the wind is correspondingly lower in the case with r-process heating. 

The r-process boosts the temperature of the early wind. At later times, most of the energy has been converted to kinetic, and the thermal energy increases due to r-process heating are modest. While the thermal energy flux is larger in the model with r-process heating, the energy density, and therefore temperature, is lower due to the much higher velocity of the material.

By construction, the mass injection rate $\dot M$ is the same in both GRHD simulations. $\dot M$ and its integral are shown in Fig.~\ref{harmsedona:fig:mdot_in}. The bulk of the mass enters the grid in the first $2-3$ seconds. 

We use the boundary data calculated from F19 for the duration of their GRMHD simulation ($\sim9\,\mathrm{s}$). We interpolate the boundary conditions linearly in time and meridional angle. After $9\,\mathrm{s}$ have elapsed, we linearly taper the velocity at the boundary to zero, while setting the other variables equal to the floor values.

\subsection{GRHD simulations in 2D}
\label{harmsedona:sec:method_harm}

We use HARMPI\footnote{Available at \url{https://github.com/atchekho/harmpi}} \citep{tchekhovskoy:19}, a parallel version of the code HARM \citep{gammie:03, noble:06}, to perform 2D axisymmetric GRHD simulations of the long-term disc wind evolution. We use the same Kerr metric as in F19, though the spacetime is approximately flat at the radii of interest. We work in modified spherical-polar Kerr-Schild coordinates. Our domain extends from $r_b=4552 r_g$ to $r_\mathrm{out} = 10^6 r_b$, and is discretized into 1024 radial points. The first 895 are spaced logarithmically between $r_b$ and $r_t = 10^4 r_b$. The remaining 129 are spaced progressively more sparsely between $r_t$ and $r_\mathrm{out}$. This would allow us to run our simulation until $t \approx r_t / c = 680\,\mathrm{s}$ before encountering any artifacts due to the grid boundary. The meridional grid is the same as in F19, but with double the number of cells. It covers the interval $[0,\pi]$, and consists of 512 cells, with resolution concentrated at the poles and near the equator. We employ an outflow boundary condition at $r_\mathrm{out}$, and a reflective boundary condition in the meridional direction. The inner radial boundary condition is a time-dependent condition, as described in the previous section. In this phase, we perform two simulations, one that uses the axisymmetrized boundary conditions, and another that uses the r-process adjusted boundary conditions.

Throughout, we use an ideal gas equation of state with fixed adiabatic index $\gamma_\mathrm{ad} =4/3$. Our simulations include the (anti)neutrino emission and alpha particle recombination models detailed in F19. However, the temperatures in our simulation are below the thresholds where these processes are significant, so the composition of the flow is a passive scalar and does not affect our results.

In the simulation that uses the r-process adjusted boundary conditions, we include an r-process heating term. During each time step $\Delta t$,
\begin{equation}
    \Delta \varepsilon_\mathrm{rad} = \rho \eta(t) \frac{\Delta t}{u^t},
\end{equation}
is added to the fluid-frame internal energy in between updates to the conserved quantities. While we evaluate the heating rate per mass $\eta(t)$ (equation \ref{harmsedona:eq:erad}) in the lab frame, we still compute the added energy in the fluid frame. The factor of $u^t$ arises because the fluid-frame time step is $\Delta t / u^t$.

We evolve our simulations to at least $t_H = 60\,\mathrm{s}$, when the flow has largely reached free expansion, and the structure has stopped evolving. At this point, regions of the ejecta with high Mach number suffer from numerical errors that artificially increase the internal energy. These are most prominent in the simulation with no r-process heating, whose thermal structure is not of  interest. When constructing the \textsc{Sedona}
input model from the 2D HARM simulation with no r-process heating, we re-set the thermal energy density (but not the kinetic energy) to be that expected from radioactive decays plus adiabatic degradation,
\begin{equation}
    \varepsilon(t_H) = \frac{\rho(t_H)}{t_H} \int_{t_0}^{t_H} \eta(t') t' \d t'.
\end{equation}

\subsection{Radiation Transport Initial Conditions}
\label{harmsedona:sec:method_sedona_ic}

When the HARM models reach homology, the ejecta are still very optically thick. It is impractical to start radiation transport calculations on timescales of $\sim 60\,\mathrm{s}$ due to the computational expense. Because we focus on accurately modeling the emission on timescales of $> 1\,\mathrm{h}$, it is unnecessary to start radiation transport calculations before $t_S = 300\,\mathrm{s}$. Between $t_H$ and $t_S$, evolution is well-modeled by homologous expansion. At the start of the \textsc{Sedona} calculation, the density will be
\begin{equation}
    \rho(t_S) = \rho(t_H) \left(\frac{t_H}{t_S}\right)^3.
\end{equation}
R-process heating and adiabatic degradation of internal energy will continue as well, giving
\begin{equation}
\varepsilon(t_S) = \varepsilon(t_H) \frac{t_H}{t_S} + \frac{\rho(t_S)}{t_S} \int_{t_H}^{t_S} \eta(t')t'\d{t'}.
\end{equation}

When constructing the input models for \textsc{Sedona}, we exclude the high Mach-number regions that are susceptible to numerical errors in the internal energy. We apply density and radial cuts to the ejecta, only including cells where
\begin{equation}
    \label{harmsedona:eq:rhocut}
    \rho > \rho_\mathrm{cut} = 6.9\times 10^{-7} \left(\frac{t}{60\,\mathrm{s}}\right)^3 \mathrm{g}\,\mathrm{cm}^{-3}
\end{equation}
and
\begin{equation}
    \label{harmsedona:eq:rcut}
    \frac{r}{t} < 0.8c.
\end{equation}
This contour is overplotted on density maps in Fig.~\ref{harmsedona:fig:harm_snapshot}. From 30 seconds onwards, this contour includes approximately $95$ per cent of the mass on the grid in both simulations. Neither of these cuts affect the shapes of the light curves. Expanding the $r/t$ cut to $0.9c$ does not affect the light curves. Lowering the value of $\rho_\mathrm{cut}$ by a factor of 10 increases the mass enclosed to 98 per cent and causes the overall luminosity of the early light curve to increase slightly (by 10 per cent).

\subsection{Radiation Transport Simulations}
\label{harmsedona:sec:method_sedona}
To predict bolometric and broad-band light curves from these disc winds, we use \textsc{Sedona}, a time-dependent, multi-dimensional, multi-wavelength Monte Carlo radiation transport code \citep{kasen:06,roth:15}. \textsc{Sedona} tracks the emission and propagation of packets of radiation (`photons') through the freely-expanding ejecta. 
Interactions between the particles and fluid are calculated in the fluid frame, which naturally accounts for adiabatic losses as well as most relativistic effects. The one relativistic effect we neglect is in the evaluation of $\eta(t)$, where we do not distinguish between proper and lab-frame time (see Section \ref{harmsedona:sec:method_harm_ic}). When particles leave the grid, they are tallied according to their wavelength, direction, and observer arrival time, providing time- and viewing-angle-dependent light curves and spectra.

The ejecta density and temperature are discretized on a cylindrical velocity grid. Since \textsc{Sedona} requires homologous expansion, we reset the velocity from the HARM models to be purely radial and have magnitude $v = r/t$. The regions within our radial and density cuts are already mostly homologous so this should not significantly affect our results. Within the \textsc{Sedona} run, density evolves as $t^{-3}$. The temperature is re-calculated at each time-step by equating the energy from photon absorption and radioactive heating with thermal emission. Adiabatic losses arise from the frame-shifting in the particle-fluid interactions. 

The initial radiation field is represented by $10^7$ particles. At each time step, $3\times 10^5$ photons are created, representing the emission from r-process decays. We use a parametrized,
time-dependent r-process heating rate that assumes an initial entropy of
$32\,\kb\,\mathrm{baryon}^{-1}$,
an expansion time-scale of $0.84\,\mathrm{ms}$, and electron fraction of $0.13$ \citep{lippuner:15}. This rate is adjusted by a time-dependent thermalization fraction that ranges between $\sim 50-75$ per cent \citep{barnes:16}. 
The resultant net heating rate roughly agrees with equation (\ref{harmsedona:eq:erad}) between $t_H$ and $t_S$. As in the HARM calculation, we use the lab frame time to evaluate $\eta$.
This may have a larger effect in the light curve calculation since the instantaneous luminosity is more important than the total energy budget. \textsc{Sedona} uses a flat spacetime, so the time dilation factor is $\gamma$. For a power-law heating rate $\propto t^{-a}$, we underestimate the heating rate by a factor of $\gamma^a$, and the total deposited energy by $\gamma^{a-1}$.
The fastest material on the \textsc{Sedona} grid has $\gamma = 1.7$.
After $t_S$, $a \approx 1.3$, so for the fastest material, we may underestimate the r-process luminosity by a factor of 2, and the total energy deposited within a time-step by $20$ per cent.
We expect the overall magnitude of this effect to be small since 98 per cent of the mass in the \textsc{Sedona} simulation has $v<0.5c$, for which luminosity is underestimated by less than 20 per cent.

We adopt a uniform grey (frequency-independent) opacity $\kappa = 1\cmspg$. This is approximately the Planck mean opacity of a mixture of first-peak r-process elements at 1 day post-merger \citep{kasen:13, tanaka:20}. On a similar time-scale, a mixture of Lanthanides has a higher average opacity, $\sim 10\cmspg$. We are primarily interested in the relative differences between ejecta with and without dynamic r-process, so the choice of a particular grey opacity will not affect the comparison, since the models including and excluding dynamic r-process heating will be affected in the same way.  

\section{Hydrodynamic results}
\label{harmsedona:sec:result_harm}

\begin{figure}
    \centering
    \includegraphics{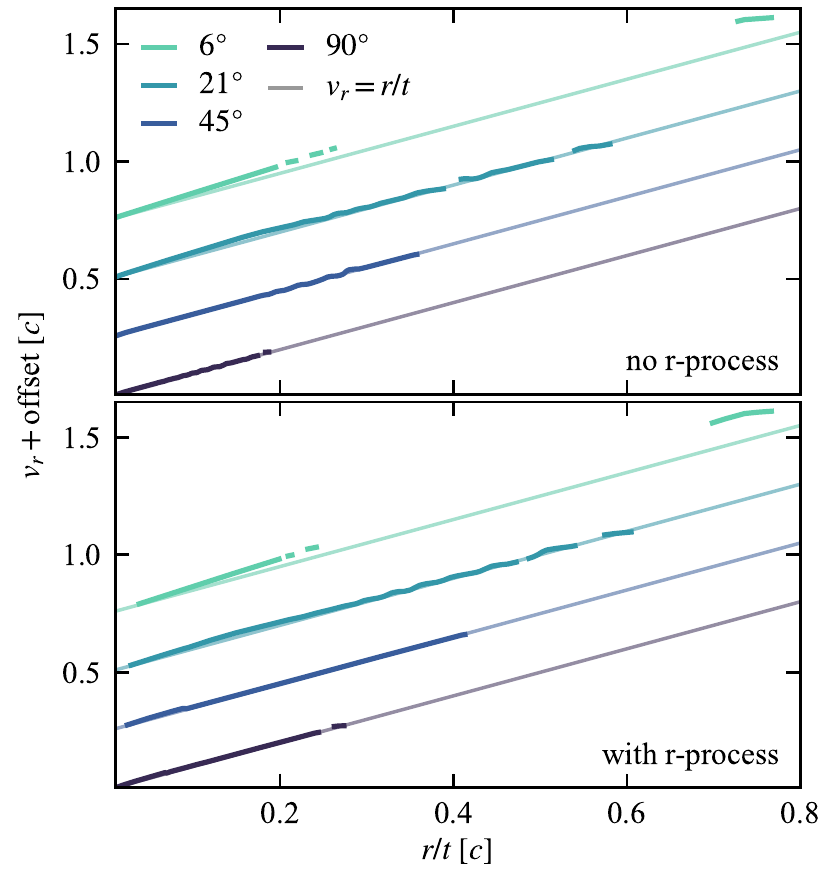}
    \caption{Radial profiles of radial velocity at a set of representative angles at $60\,\mathrm{s}$ (dark colors) in our long-term 2D GRHD simulations of NS merger disc winds. Angles are measured from the South pole. Different angles are separated by an offset. The light colored lines show homology ($v_r = r/t$) for a given angle, which is generally an excellent approximation to the simulation results; most of the motion is self-similar. The profiles are only shown where $\rho > \rho_\mathrm{cut}$ (equation \ref{harmsedona:eq:rhocut}). The top (bottom) panel shows results from the simulations without (with) r-process heating. }
    \label{harmsedona:fig:homology}
\end{figure}

We follow our 2D GRHD simulations to 60 seconds, by which point the flow is mostly in homology ($v \propto r$). Fig.~\ref{harmsedona:fig:homology} compares velocity profiles for selected angles against the profile that \textsc{Sedona} imposes on the ejecta ($v = r/t$). Deviations from homology are small, especially in the lower-velocity regions that contain most of the mass. Further evolution in GRHD brings the polar regions in line with the velocity structure assumed by \textsc{Sedona}, but allows for more numerical errors to accumulate in the internal energy in high Mach-number regions. After 60 seconds, changes to the light curves due to these errors exceed those from further evolution of the density structure.

\begin{figure}
  \includegraphics{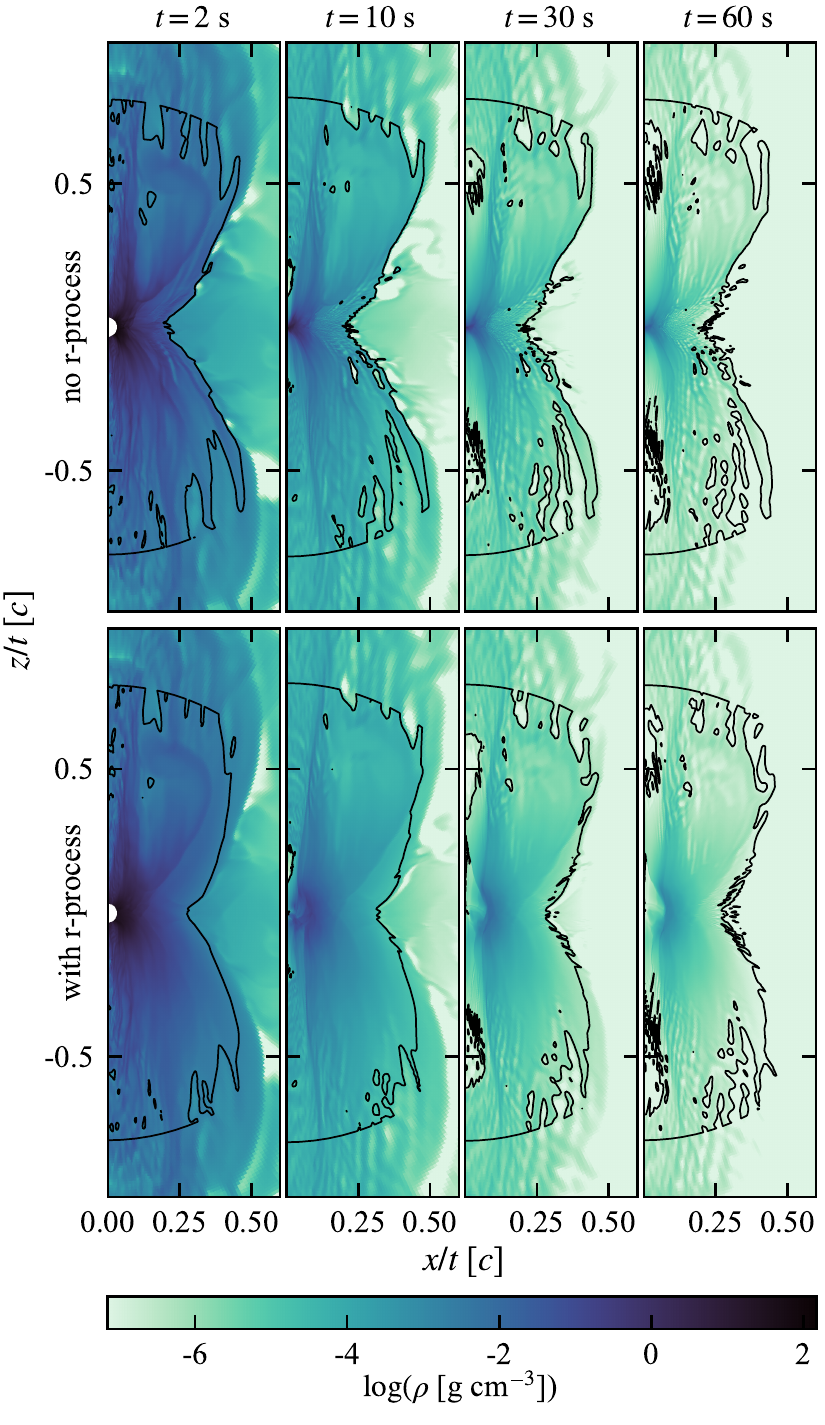}
  \caption{Snapshots of mass density after 2, 10, 30, and 60 seconds of evolution in the long-term 2D GRHD simulations discussed in section \ref{harmsedona:sec:method_harm}. The top (bottom) panels show the runs without (with) r-process heating. The spatial bounds are scaled by the snapshot time. The black contour bounds the region included in the \textsc{Sedona} radiation transport simulations ($r/t<0.8c$ and $\rho > \rho_\mathrm{cut}$). The primary consequence of including r-process heating is that the low-velocity portion of the wind is accelerated to $\sim 0.1c$. Heating also smooths out the finer density structure. The effects also become more prominent as time progresses, and the gas thermal energy has time to be converted to kinetic energy.} 
    \label{harmsedona:fig:harm_snapshot}
\end{figure}

Density snapshots from our 2D GRHD simulations are shown in Fig.~\ref{harmsedona:fig:harm_snapshot}. In the first few seconds, there are minimal differences between the two simulations. The r-process energy has not yet caused the gas to expand or accelerate. By 10 seconds, the effects become noticeable, as the heating blurs some of the small-scale structure in the wind. The broad structure stabilizes around 30 to 60 seconds. R-process heating accelerates the slowest material to $\sim 0.1c$, which in this case, evacuates a sphere of radius $0.1ct$ at the center of the domain.

\begin{figure}
  \includegraphics{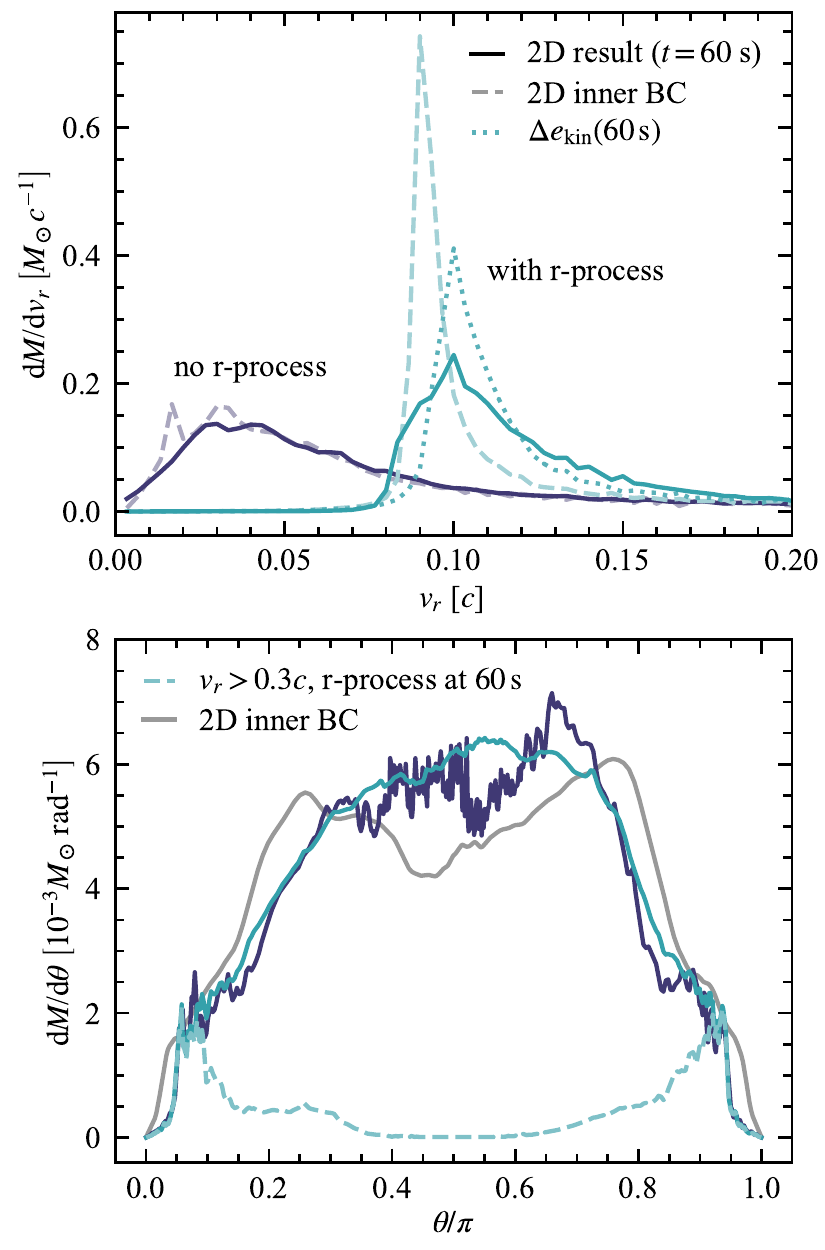}
  \caption{Distribution in $v_r$ (top panel) and meridional angle $\theta$ (bottom panel) of the material at the end of the 2D GRHD simulations ($t=60\,\mathrm{s}$; solid lines and darker colors). We compare the simulations with (teal) and without (purple) r-process heating. Top panel: The radial velocity distributions of material injected at the inner boundaries of the 2D simulations are shown in lighter colors and dashed lines. The dotted line shows the mass distribution obtained by applying the formalism of Section \ref{harmsedona:sec:method_harm_ic} but evaluating $\Delta e_\mathrm{kin}$ at $60\,\mathrm{s}$. Bottom panel: The angular distribution of the input is the same in both simulations and is shown in grey in the bottom panel. The distributions remain similar at 60 s, though the inclusion of r-process heating smooths small-scale structure in the distribution. The material near the pole is almost all high-velocity ($v_r>0.3c$, dashed line in bottom panel), which is not affected by r-process heating. }
    \label{harmsedona:fig:dmdb_dmdh}
\end{figure}

\begin{figure}
    \centering
    \includegraphics{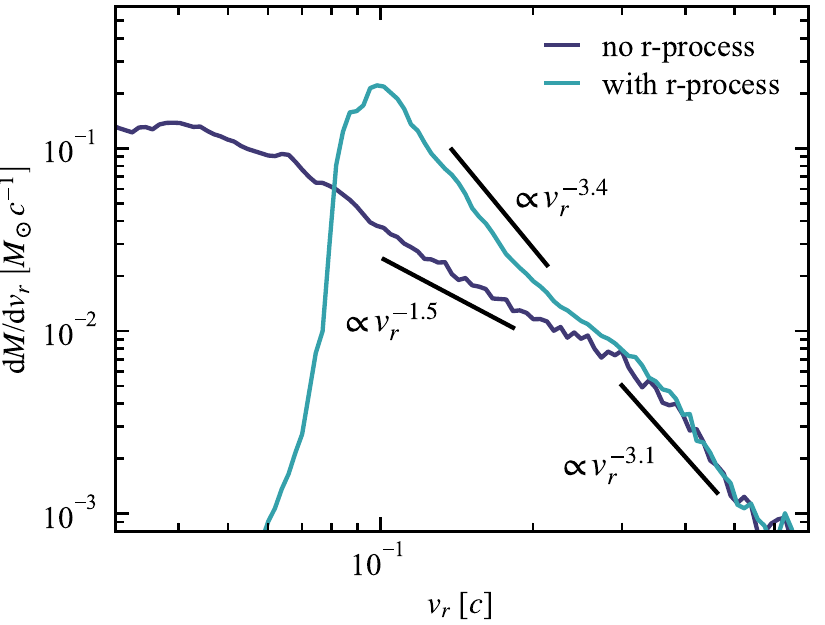}
    \caption{Same as the top panel of Fig.~\ref{harmsedona:fig:dmdb_dmdh}, but showing the high-velocity tail of the mass distribution, and only showing the resulting distributions at $60\,\mathrm{s}$. The slopes of power-law fits to $\mathrm{d}M/\mathrm{d}v_r$ for both models are shown in black. We separately fit low-velocity $\lesssim 0.3c$ material and high-velocity matter, since the latter is dynamically unaffected by r-process heating.}
    \label{harmsedona:fig:dmdb_log}
\end{figure}

The initial and final $\mathrm{d}M/\mathrm{d}v_r$ and $\mathrm{d}M/\mathrm{d}\theta$ distributions are shown in Fig.~\ref{harmsedona:fig:dmdb_dmdh}. In both models, the vast majority of the mass is at relatively low velocity $\lesssim 0.2 c$. Without r-process heating, the velocity distribution of the wind is mostly  unchanged by further hydrodynamic evolution. There is much more of a change between the initial and final distributions with the inclusion of r-process heating. The light teal line in the top panel of Fig.~\ref{harmsedona:fig:dmdb_dmdh} is the inner boundary condition in our GRHD simulation with r-process heating. This shows a distribution that is much more sharply concentrated in velocity than the final configuration after the subsequent r-process heating and hydrodynamic evolution (dark teal line). The input distribution has almost no material with $v>0.15c$. That is because the higher-velocity material is the first to enter the domain and has not yet a) experienced as much r-process heating, b) converted it to kinetic energy. A check of the formalism of Section \ref{harmsedona:sec:method_harm_ic} (equations \ref{harmsedona:eq:energy_tot}-\ref{harmsedona:eq:quadenergy}) is to evaluate $\Delta e_\mathrm{kin}$ at $60\,\mathrm{s}$, and compare the resulting distribution to the results of the 2D GRHD simulation at $60\,\mathrm{s}$. The formalism accurately predicts the modal velocity of the wind, but underestimates the spread in the radial velocity distribution. The difference is especially pronounced for velocities below the mode ($v \lesssim 0.1c$). 

It is also possible that our GRHD simulations with r-process heating underestimate the amount of mass with $v<0.07$. In the underlying 3D GRMHD simulation, there is marginally-bound material that reaches a maximum radius $<r_b$, but would become unbound with the additional boost from r-process heating. The radial cut at $r_b$ excludes this material, in effect applying a total energy floor to the ejecta. This enforces an artificial minimum velocity on the wind, which could in turn suppress the central density in our r-process model. Ideally, r-process heating should be included from the start of the GRMHD simulations of NS merger discs.

As expected, $\mathrm{d}M/\mathrm{d}v_r$ is unaffected by r-process heating above a velocity of $0.3c$ (Fig.~\ref{harmsedona:fig:dmdb_log}). In that region, we find that $\mathrm{d}M/\mathrm{d}v_r \propto v_r^{-3.1}$. We exclude material where $\rho < \rho_\mathrm{cut}$, but it is still possible that the slope of the mass distribution is affected by the density floor of the 3D GRMHD simulation, so this fit may not be that accurate. Below $0.3c$, r-process heating causes the slope to steepen, giving a rough power law fit $\mathrm{d}M/\mathrm{d}v_r \propto v_r^{-3.4}$, as compared to the distribution $\mathrm{d}M/\mathrm{d} v_r \propto v_r^{-1.5}$ found when fitting the distribution in the simulation without r-process heating. Fits are calculated over the interval $0.1c \leq v_r \leq 0.25c$ for the model with r-process heating, and $0.03c\leq v_r\leq 0.25c$ for the model without. The values of the slopes are unlikely to be that robust, but the steepening of the slope is a general prediction of r-process heating.

The angular distributions are also similar between the GRHD simulations with and without r-process. In both cases, the mass distribution is more equatorially-concentrated at the end of the 2D HARM simulations than it is at injection. This is due to the meridional component of the velocity at injection. The inclusion of r-process heating smooths some of the features in angular structure, and slightly broadens the structure. The poles remain relatively evacuated, which may be artificial, since 87 per cent of the energy from r-process is deposited off of the grid, i.e.\ prior to injection into the 2D GRHD simulation. 
The large fraction of the energy injected off-grid may lead our models to underestimate the extent to which the lower-velocity material will fill in the jet cavity. 
We account for the conversion of thermal to kinetic energy, but only apply this in the direction of the existing velocity vector. Our formalism does not account for meridional expansion that could occur from the early heating prior to injection.

That said, almost all of the polar material is moving faster than $0.3c$ (Fig.~\ref{harmsedona:fig:dmdb_dmdh}), which is a portion of the wind whose kinematics seem mostly unaffected by r-process heating. We also neglect magnetic fields, which will likely resist the gas's expansion into the cavity at the pole.

\section{Radiation transport results}
\label{harmsedona:sec:result_sedona}

\begin{figure}
    \includegraphics{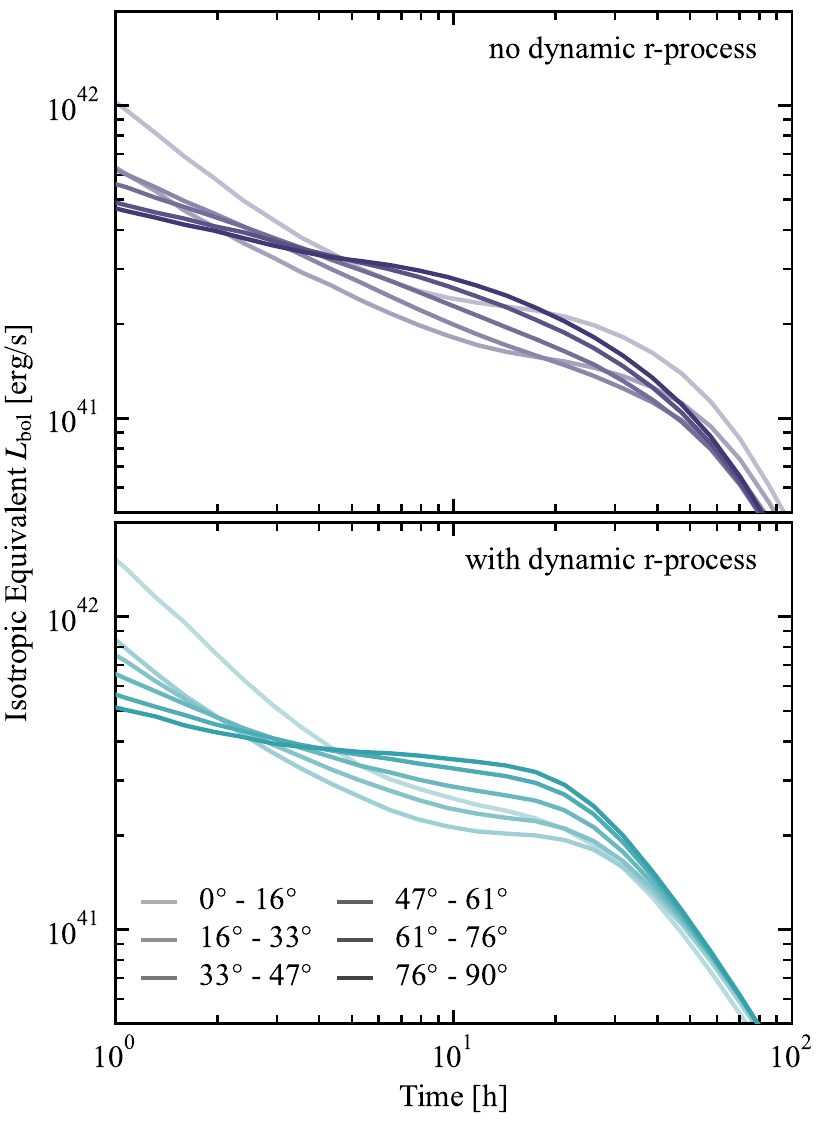}
    \caption{Viewing angle-dependent isotropic equivalent bolometric light curves for models without (top) and with (bottom) r-process heating during the hydrodynamic calculations. Both models include r-process heating in the light curve calculation. The light curves shown are averages for observers within the given angular ranges. Polar angles are shown in lighter colors, and equatorial angles in darker ones. For clarity, only Southern hemisphere viewing angles are shown; angles are measured from the South pole.
    Polar angles are brighter in the first few hours, after which equatorial viewing angles are brighter by a factor of $\sim 2$. The latter is likely due to the prolate wind structure.}
    \label{harmsedona:fig:lc_bol}
\end{figure}

\begin{figure*}
    \includegraphics{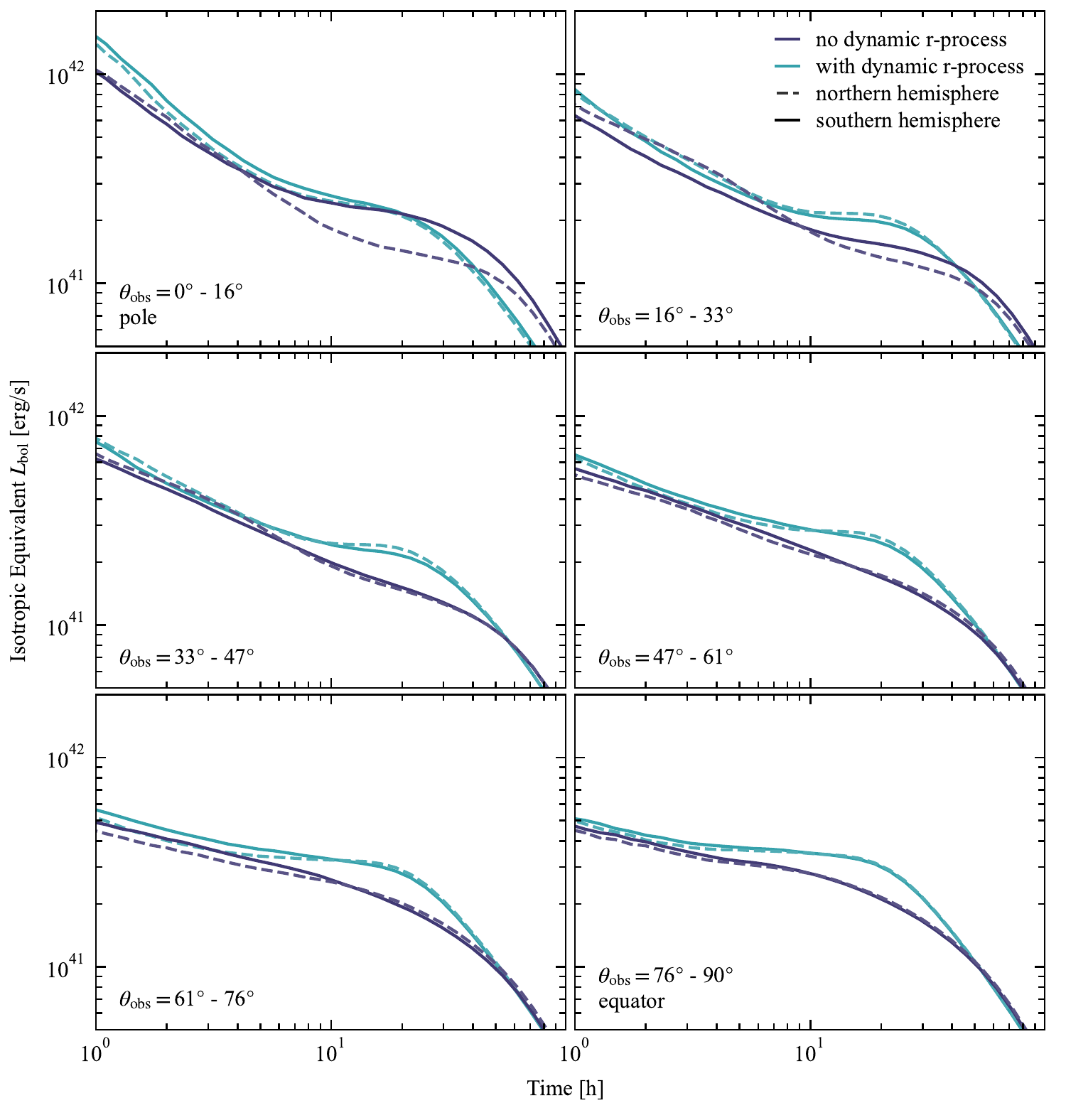}
    \caption{Isotropic equivalent bolometric luminosity light curves, as in Fig.~\ref{harmsedona:fig:lc_bol}, but each panel shows a different viewing angle range. Light curves from the model with (without) dynamic r-process are shown in teal (purple). Southern hemisphere light curves ($\theta_S = 180\dg - \theta_N$) are shown as solid lines; Northern hemisphere light curves are dashed. The main effect of dynamic r-process heating is to brighten and flatten the light curve from $0.3-3\,\mathrm{d}$. The light curve also declines somewhat more quickly, most noticeably near the poles. This is due to the change in characteristic velocity from $\sim 0.03c$ to $\sim 0.1c$. The high-velocity material is minimally affected by r-process heating, so the early time portion of the light curve is nearly the same. The South pole is brighter than the North when there is no dynamical r-process heating, likely due to a non-general feature in the 3D GRMHD simulation. Dynamical r-process heating eliminates this difference.}
    \label{harmsedona:fig:lc_bol_trans}
\end{figure*}

Using the Monte Carlo radiation transport code \textsc{Sedona}, we calculate viewing angle-dependent bolometric and band light curves for our disc wind models with and without dynamical r-process heating. We include r-process heating in both \textsc{Sedona} simulations, since instantaneous r-process heating at time $\sim t$ sets the luminosity at time $\sim t$. The code assumes homologous expansion, which is a good approximation at this stage; the heating does not affect the dynamics and only powers the kilonovae.

Our results are shown in Figs~\ref{harmsedona:fig:lc_bol}~and~\ref{harmsedona:fig:lc_bol_trans}. In both models, polar viewing angles are brighter in the first $2-3\,\mathrm{h}$. Afterwards, the equator brightens relative to the polar regions. While there is not much high-velocity mass, it is concentrated around $\sim 15\dg$ from the poles, and is optically thick in the first few hours of evolution. Photons from a fast-moving photosphere near the pole are Doppler shifted towards the observer, resulting in particularly bright early polar emission. Due to the low density directly on the poles, polar observers are also able to see `deeper' into the ejecta, which can lead them to see a hotter and brighter photosphere on the poles, even if said photosphere is at a lower velocity \citep{klion:21}. 

The slope of the high-velocity mass distribution, especially at the pole, may be sensitive to detailed physics not included in our calculations (e.g.\ equation of state, neutrino transport, and the exact magnetic field structure in the remnant disc). However, the general correlation between high-velocity material and bright, early light curves is plausibly robust. At later times, $\gtrsim 1\,\mathrm{d}$, the polar light curves continue to fall more sharply, while the equatorial light curves remain flat. Between 4 hours and 1 day, the equatorial light curves are brighter than on the poles due to the prolate structure of the wind. The greater projected surface area on the equator causes equatorial light curves to be somewhat brighter \citep{darbha:20, korobkin:21}. 

The effects of dynamical r-process heating are most clearly seen when comparing light curves from the same viewing angles (Fig.~\ref{harmsedona:fig:lc_bol_trans}). We also compare light curves in the Northern and Southern hemispheres, which gives an estimate of the minimum uncertainty in the light curves. The setup of the 3D GRHMD simulation is equatorially symmetric, so on average, the light curves from the North and South should be the same. However, in our calculations the South pole is substantially brighter than the North when there is no r-process heating in the axisymmetrized GRHD run. This is likely due to a numerical artefact on the North pole of the 3D GRMHD simulation that disrupts the polar density structure. The South pole light curves are likely somewhat more reliable than those on the North pole. However, the details of the polar structure of the ejecta are, like the high-velocity distribution, sensitive to physics details that we do not explore here. The difference between the hemispheres goes away once r-process heating is included, which suggests that r-process heating may homogenize the wind structure. 

The early light curve is unaffected by r-process heating, consistent with the interpretation that it primarily arises from high-velocity material that is too fast to be substantially affected by r-process heating. The primary effect of dynamical r-process heating is to make the light curves brighter but fade more quickly. This is consistent with the scaling between light curve time-scale and ejecta velocity, $t_\mathrm{peak} \propto v^{-1/2}$ \citep{metzger:19}. On the South pole, the light curves from the models with and without dynamic r-process heating are quite similar from $\sim 6-30\,\mathrm{h}$. This may be a coincidence. 

Our model with dynamic r-process heating has a bolometric luminosity of $\sim 3\times 10^{41}\,\mathrm{erg/s}$ at 1 day. We do not directly compare our models with observations of AT 2017gfo since our wind has a mass of only $0.013\msun$, differing by at least a factor of 3 from the $>0.04\msun$ of AT 2017gfo \citep[e.g.][]{villar:17}. The peak luminosity for a kilonova is expected to scale as $L_\mathrm{peak} \propto M^{0.35}$, while the peak time (and therefore time-scale) $t_\mathrm{peak} \propto M^{1/2}$ \citep{metzger:19}. Applying this correction brings our light curves to the rough energy and time-scale of AT 2017gfo ($4\times 10^{41} \,\mathrm{erg/s}$ at $1.7\,\mathrm{days}$), though the light curves have a different shape before $\lesssim 1\,\mathrm{d}$.
There is somewhat better agreement between the observations and the model with r-process heating. That is likely because the characteristic velocity of our r-process wind is higher, more consistent with the inferred velocity of AT 2017gfo.

\begin{figure*}
    \includegraphics{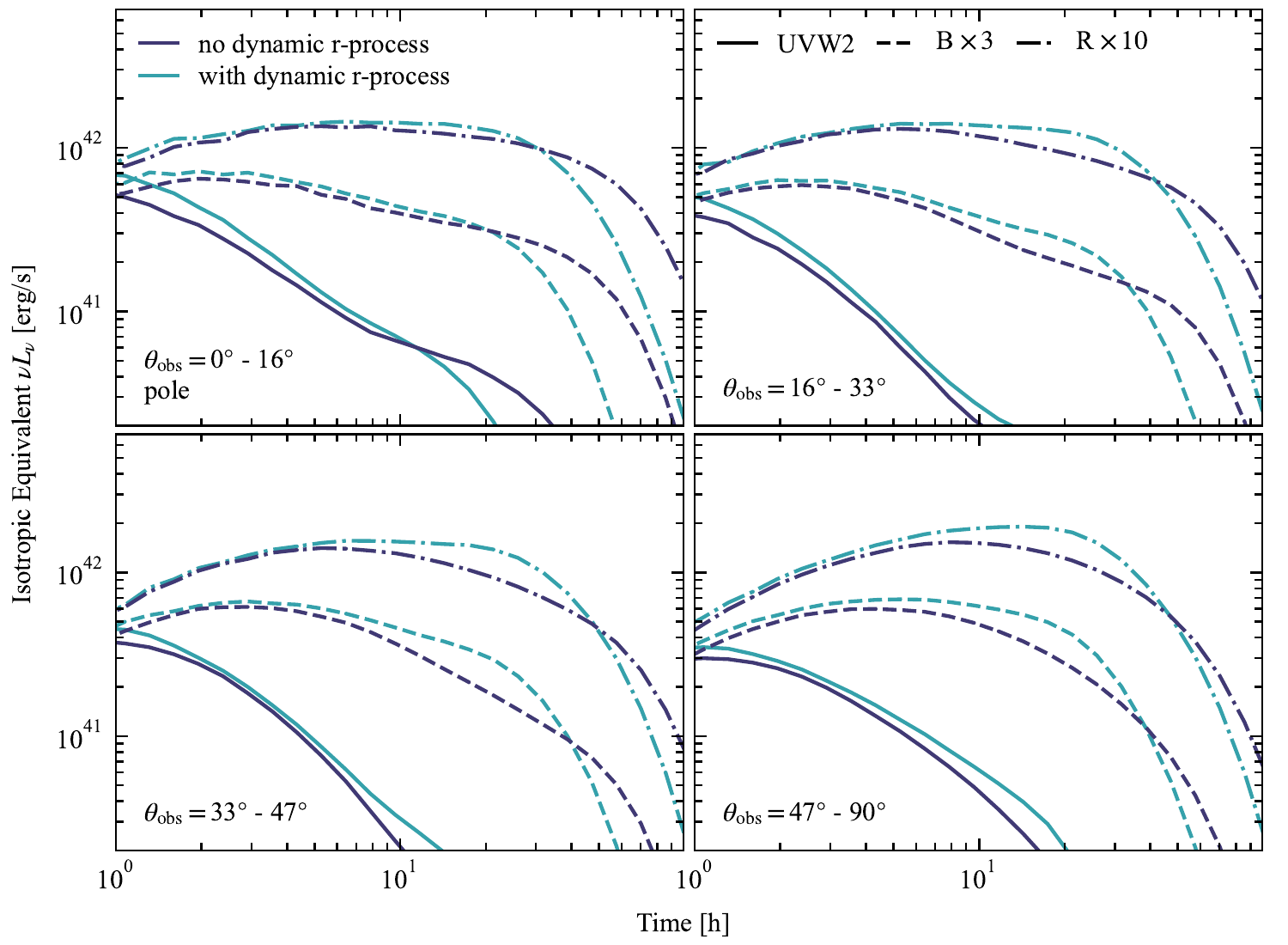}
    \caption{\emph{Swift} UVOT W2 (solid), B (dashed), and Cousins R-band (dashed-dotted) light curves for Southern hemisphere viewing angle ranges (measured from the South pole). Viewing angles $>47\dg$ are similar and are therefore shown together. Models without (with) dynamic r-process heating are in purple (teal). As with the bolometric light curves, dynamic r-process only affects the light curves after $8$ hours. When dynamic r-process is included, the light curves are brighter and fade more quickly. Equatorial B and R-band light curves peak later than polar viewing angles, which have flatter light curves in the redder bands.}
    \label{harmsedona:fig:lc_band}
\end{figure*}

The broadband light curves follow similar trends to the bolometric. We show \emph{Swift} UVOT W2 (average wavelength 193 nm), Johnson B (442 nm), and Cousins R-band (635 nm) light curves in Fig.~\ref{harmsedona:fig:lc_band}. Across all bands, the polar angles are brighter than equatorial ones in the first few hours. Subsequently, the equatorial light curves brighten, while the emission near the pole remains constant or starts to fade. On time-scales of a day, equatorial emission is both brighter and bluer than the poles. Similar to the bolometric light curves, the dynamical effects of r-process heating are only apparent at later times when the photosphere reaches the slower (and therefore more affected) velocities. The difference in the evolution time-scales between models with and without r-process heating is more apparent in the band light curves, especially near the pole. The colors predicted here could also change noticeably with the inclusion of more realistic r-process opacities. Line blanketing, particularly in the blue and ultraviolet, is expected. This would make the UVOT W2 and $B$-band light curves dimmer, while causing more re-processed emission in the red and infrared. However, opacities at early ($\lesssim 1\,\mathrm{d}$) phases of kilonova evolution may be lower than they are later \citep{banerjee:20}. The effects of realistic opacities on the colours of these models will be an interesting direction for future work.

\begin{figure}
    \centering
    \includegraphics{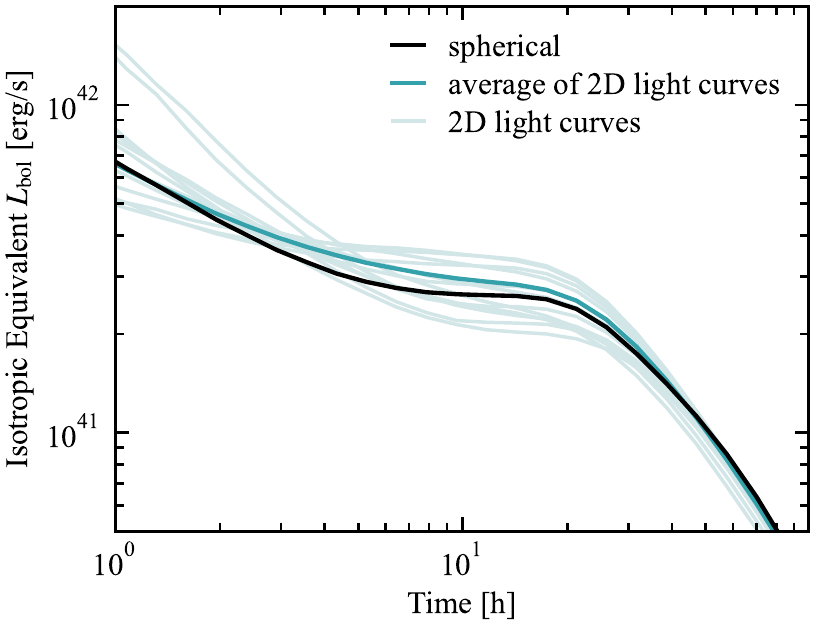}
    \caption{Comparison of spherically averaged light curves for the model with dynamic r-process heating. The black line shows the light curve obtained by spherically averaging the \textsc{Sedona} model before the radiation transport. At a few hours, this is 20 per cent dimmer than the average of the 2D light curves shown in Fig.~\ref{harmsedona:fig:lc_bol}. Light curves from individual viewing angles are shown in light teal, while the average is in dark teal. }
    \label{harmsedona:fig:lc_spher}
\end{figure}

If we spherically average the ejecta before the \textsc{Sedona} calculation, we find that the light curve obtained is similar to the average of all viewing angles in the 2D calculation. We make the comparison between the spherical, averaged, and individual 2D light curves in Fig.~\ref{harmsedona:fig:lc_spher}. For most viewing angles, the light curves have a similar shape, mostly unaffected by the asymmetry of the ejecta. Off of the poles, the light curve is well-modeled by a spherical average. The asymmetry only matters for the polar viewing angles.

\section{Discussion \& Conclusion}
\label{harmsedona:sec:discussion}

The kinematic effects of r-process heating on a kilonova disc wind can be substantial, accelerating the ejecta from a mass-weighted median velocity of $0.06c$ to $0.12c$ (Fig.~\ref{harmsedona:fig:dmdb_dmdh}). The faster wind leads to a transient that evolves more quickly and is brighter and bluer on $\sim \mathrm{day}$ time-scales (Figs~\ref{harmsedona:fig:lc_bol}~and~\ref{harmsedona:fig:lc_band}). A factor-of-two increase in ejecta velocity will lead to a transient that evolves $1.4$ times as quickly. Quantitative predictions of kilonova light curves from disc wind simulations should therefore account for the kinematic effects of r-process heating. The wind velocities from these simulations are sensitive to the initial magnetic field configuration. A toroidal configuration and/or a weaker field than the one used in F19 would produce a slower wind, which would experience a larger change in velocity when heated by the r-process.

The early ($\lesssim 4\,\mathrm{h}$) kilonova light curve is primarily produced by the high-velocity ($\gtrsim 0.3c$) ejecta, which is concentrated near the poles. The poles are especially bright early on because the photosphere is at high velocity, and the emission is Doppler shifted towards the polar observers \citep{kasliwal:17}. The high-velocity portion of the ejecta is mostly unaffected by r-process heating, so the early light curves are similar to one another in the models with and without dynamic r-process heating. This portion of the light curve is likely sensitive to the physics in the underlying 3D GRMHD simulation (e.g.\ neutrino interactions, realistic magnetic field structure, and equation of state). Additionally, the high-velocity tail is likely to be disrupted as it runs into the dynamical ejecta from the merger. Joint modeling of the dynamical and secular ejecta is likely necessary to predict the early light curves.

After the first few hours, equatorial viewing angles become brighter than polar ones due to the prolate structure of the ejecta, which has a larger projected surface area at the equator. As the photosphere recedes into the slower portions of the ejecta, the kinematic effects of r-process heating become apparent. The light curves from the model with dynamical r-process heating are flatter at $\sim 0.5-1.5\,\mathrm{days}$, and also fade earlier. The faster evolution is particularly apparent near the poles.

The difference we observe in the low-velocity distributions with and without r-process heating contrasts with the results of \citet{kawaguchi:21}. They find that the effects of including r-process heating during the hydrodynamic phase of the calculation are very small. As they discuss, this is likely because they do not account for heating that occurs before the wind enters their simulation grid. In their simulations, most material enters at $\gtrsim 2\,\mathrm{s}$, which is longer than the $1\,\mathrm{s}$ time-scale on which most of the r-process heating occurs. In our calculations, we modify the boundary conditions of the GRHD simulation to account for r-process heating that occurred before injection, thus capturing in an approximate way the dominant dynamical effects of r-process heating.

Our formalism for pre-injection heating roughly captures the conversion of r-process heating to kinetic energy. However, we are unable to include all of the effects of the early heating. In particular, we do not account for the meridional expansion of the gas, which may be significant, particularly at the poles. The propagation of the jet leaves a polar cavity that may be filled by hot material from the wind. In our simulation with r-process heating, the evacuated region survives. It is possible that the self-consistent inclusion of r-process heating at all times would cause the gas to expand more and fill in the polar region. There is also the competing effect of magnetic fields, which resist polar expansion of the ejecta. A 3D GRMHD accretion disc simulation that includes r-process heating is necessary to understand the interaction between magnetic fields and hot gas in the polar region.

The low-velocity distribution in the 2D GRHD simulation with r-process heating is truncated at $0.07c$, which may not be physical. In the simulation of F19, there is material that is marginally bound and reaches a maximum radius $< r_b = 2\times 10^4\,\mathrm{km}$. With r-process heating, some of this material would have become unbound and crossed $r_b$. Because we apply a cut at $r_b$, this material is excluded, even though it would have formed part of the wind. This may artificially truncate the velocity distribution, making the late-time light curve fall off more quickly than it would otherwise.

The degree to which the low-velocity distribution is truncated will also have implications for kilonova spectra. \citet{kasen:15} found that slowly-moving ($\sim 0.03c$) ejecta have resolved absorption lines that could possibly be used to identify elements in kilonova ejecta. Faster ($0.1-0.3c$) ejecta have broadened lines that make such identification more difficult \citep{kasen:13}.

This underscores the importance of self-consistently including r-process heating within the accretion disc simulations themselves. Accurately capturing the first few seconds of heating and their kinematic effects is critical for predicting both the radial velocity and angular distributions of the wind.

In our radiation transport calculations, we adopt a grey opacity. Kilonova ejecta are composed of dozens of elements with large, highly frequency-dependent opacities. Our qualitative light curve results are likely robust to changes to the opacity (e.g.\ faster ejecta leading to faster light curve evolution). The effects of more realistic opacities are none the less an important direction for future investigations. The majority of r-process products have particularly high opacities in the ultraviolet and blue portions of the spectrum, so our grey prescription may lead us to overestimate the UV and blue emission. However, at very early times ($<0.5\,\mathrm{day}$), the opacity may be suppressed due to the high temperatures and ionization fractions \citep{banerjee:20, klion:21}. 

The composition of disc winds is expected to vary, depending on a number of factors including the life time of a central (hyper-massive) neutron star \citep{lippuner:17}. A single event can produce material with a wide range of $\ye$, and the distribution can be both radially and meridionally-stratified \citep[e.g.][]{just:15, siegel:17, fernandez:19}. The overlay of distinct ejecta components with different opacities may have interesting consequences for viewing angle-dependent kilonova light curves \citep{korobkin:21}.

\section*{Acknowledgements}
This research was funded by the Gordon and Betty Moore Foundation through grant GBMF5076. AT was supported by NASA 80NSSC18K0565 and NSF AST-1815304 grants. EQ was supported in part by a Simons Investigator award from the Simons Foundation. RF acknowledges support from the Natural Sciences and Engineering Research Council (NSERC) of Canada through Discovery Grant RGPIN-2017-04286, and from the Faculty of Science at the University of Alberta.
The simulations presented here were carried out and
processed using the National Energy Research Scientific Computing Center, a U.S. Department of Energy Office of Science User Facility, and the Savio computational cluster resource provided
by the Berkeley Research Computing program at the University of
California, Berkeley (supported by the UC Berkeley Chancellor,
Vice Chancellor of Research, and Office of the CIO).

\emph{Software:} HARM \citep{gammie:03, noble:06}, \textsc{Sedona} \citep{kasen:06, roth:15}, \texttt{matplotlib} \citep{matplotlib}, \texttt{numpy} \citep{numpy},  \texttt{scipy} \citep{scipy}, FSPS filter files \citep{conroy:09, conroy:10}.

\section*{Data Availability}

The data underlying this article will be shared on reasonable request to the corresponding author.



\bibliographystyle{mnras}
\bibliography{bibliography,non_ads} 


\bsp	
\label{lastpage}
\end{document}